\documentclass[doublespacing]{elsart}
\journal{NIM}

\usepackage{amssymb}
\usepackage{amsmath,amsfonts,verbatim,graphicx,float}
\usepackage{subfigure}
\usepackage{cite}

\newfont{\tensy}{cmsy10}

\begin{document}
 
\begin{frontmatter}



\title{Unfolding of differential energy spectra in the MAGIC experiment}

 \author[a]{J.~Albert}, 
 \author[b]{E.~Aliu}, 
 \author[c]{H.~Anderhub}, 
 \author[d]{P.~Antoranz}, 
 \author[b]{A.~Armada}, 
 \author[d]{M.~Asensio}, 
 \author[e]{C.~Baixeras}, 
 \author[d]{J.~A.~Barrio},
 \author[f]{H.~Bartko}, 
 \author[g]{D.~Bastieri}, 
% ...

 \address[a]{Universit\"at W\"urzburg, D-97074 W\"urzburg, Germany}
 \address[b]{Institut de F\'\i sica d'Altes Energies, Edifici Cn., E-08193 Bellaterra (Barcelona), Spain}
 \address[c]{ETH Zurich, CH-8093 Switzerland}
 \address[d]{Universidad Complutense, E-28040 Madrid, Spain}
 \address[e]{Universitat Aut\`onoma de Barcelona, E-08193 Bellaterra, Spain}
 \address[f]{Max-Planck-Institut f\"ur Physik, D-80805 M\"unchen, Germany}
 \address[g]{Universit\`a di Padova and INFN, I-35131 Padova, Italy} 
% ...

 \corauth[cor1]{Corresponding author.}

\begin{abstract}
The paper describes the different methods, used in the MAGIC experiment, to unfold 
experimental energy distributions of cosmic ray particles ($\gamma$-rays). Questions and 
problems related to the unfolding are discussed. Various procedures are proposed 
which can help to make the unfolding robust and reliable. The different methods and 
procedures are implemented in the MAGIC software and are used in most of the analyses. 
\end{abstract}


\end{frontmatter}

\section{Introduction}
In an Imaging-Air-Cherenkov-Telescope (IACT) experiment like MAGIC \cite{MAGIC-commissioning}
the energy $E$ of the cosmic ray particle ($\gamma$-ray)
is not exactly known. It has to be estimated, the energy
resolution being in the order of 20 to 40\%. As a consequence, the
experimentally measured energy spectrum is biased. The procedure to correct for the
effects due to the finite energy resolution is called unfolding. 

While in high-energy-physics experiments unfolding is a widely used technique, 
this is not the case in present day's IACT experiments.
This paper deals with the unfolding procedure, which is applied as a standard tool 
in the MAGIC experiment. 
The different unfolding methods are explained in detail. Emphasis is put 
on the discussion of questions and problems related to 
the application of the unfolding to real data. 
It is not the aim of the paper to give a complete derivation of all formulas. 
For this the reader is referred to the publications
\cite{gold64} to \cite{blobel02}. An excellent review of unfolding methods
is given in \cite{anykeyev91}. The present paper makes use of many ideas
discussed in that paper.

Although here only differential energy spectra are considered, the procedures 
are equally well applicable to distributions of other quantities, including 
distributions in more than one dimension \cite{wittek99}. 

The layout of the paper is as follows. In Section \ref{section:unfoldingaim} the
notation is defined and the motivation for the unfolding procedure is given. 
The different 
unfolding methods, which means the different ways of regularization, 
are presented in Section \ref{section:regularization}. The so-called
Forward Unfolding, which represents an implicit unfolding under the assumption of 
a certain parametrization of the unfolding result $S$, is explained in Section 
\ref{section:forwardunfolding}. In Section \ref{section:usefulquantities} two 
quantities are introduced, which are useful for an optimal choice of the 
regularization strength. The criteria for this choice are collected in Section
\ref{section:bestweight}. Section \ref{section:comments} discusses various 
technical aspects which are important in the application of the unfolding procedure
to real data.
Two particular technical procedures, which ensure an unbiased and robust unfolding,
are presented in Sections \ref{section:AeffandM} and \ref{section:combine}. Some 
unfolding results, obtained by applying the unfolding procedure 
to data taken in the MAGIC experiment,
are discussed in Section \ref{section:application}. Finally the pros and contras of  
the method of Correction Factors, which is an alternative way of correcting data 
for effects due to the finite experimental resolution, are listed in Section 
\ref{section:correctionfactors}. A summary is given in Section \ref{section:summary}.

\section{The aim of the unfolding procedure}
\label{section:unfoldingaim}
In this Section the notations are defined and the motivation for the unfolding procedure
is explained.

\subsection{Notation}
\label{section:notation}
The true and measured (estimated) values of the energy of the cosmic ray particle 
are denoted by $E_{true}$
and $E_{est}$ respectively. The data are assumed to be binned in histograms, and
certain binnings are chosen independently for the
distributions in  $E_{true}$ and $E_{est}$. Furthermore, the following definitions 
are introduced: \\

\begin{tabular}{cll}
 $Y_i$     & number of events in bin $i$ of $E_{est}$  & $(i=1, ... na)$ \\
 $K_{i,k}$ & covariance matrix $K$ of $Y$              & $(i=1, ... na;\;\;k=1, ... na)$ \\
 $S_j$     & number of events in bin $j$ of $E_{true}$ & $(j=1, ... nb)$ \\
 $T_{j,l}$ & covariance matrix $T$ of $S$              & $(j=1, ... nb;\;\;l=1, ... nb)$ \\
 $M_{ij}$  & migration matrix $M$            & $(i=1, ... na;\;\;j=1, ... nb)$ \\ 
 $G_{ik}$  & Gram's matrix $G=M\cdot M^T$& $(i=1, ... na;\;\;k=1, ... na)$ \\ 
\end{tabular} \\

with
\begin{alignat}{2}
\sum_{i=1}^{na}{M_{ij}}\;&=\;1\qquad 
\qquad{\rm for\;all}\;j   
\label{eq:normalizeM}
\end{alignat}

In the following it is assumed that the rank $nr$ of $G$ is equal to (and not less than) 
the minimum of $na$ and $nb$, where $na$ and $nb$ are the number of bins in $E_{est}$ and
$E_{true}$ respectively, which are used in the unfolding.  
This can always be achieved by a proper choice of the
binnings in $E_{true}$ and $E_{est}$.

The migration matrix contains the most likely fraction
of events moving from a bin $j$ in $E_{true}$
into a bin $i$ of $E_{est}$, due to the finite experimental energy resolution:
\begin{alignat}{2}
{Y}_i\;&=\;\sum_{j=1}^{nb}{M_{ij}\cdot {S}_j}\qquad(i=1, ...na) \nonumber\\
\qquad {\rm or\;in\;matrix\;notation} 
\qquad Y\;&=\;M\cdot S
\label{eq:YMS}
\end{alignat}

The migration matrix $M$ is obtained from Monte Carlo (MC) simulations, in which the 
development of the air shower (induced by the cosmic ray particle in the 
atmosphere), the emission 
of Cherenkov light in the air shower, the geometrical, optical and electronic
properties of the telescope and the experimental procedures in the data analysis
(shower reconstruction, $\gamma$/hadron separation, 
energy estimation, selections, cuts) are simulated \cite{Majumdar2005}. $M$ 
is computed from a 2-dimensional plot of the number of reconstructed MC events in the
$E_{est}$-$E_{true}$ plane,
which was produced under the same conditions (selections, cuts) 
as the experimental distribution $Y$. It is the aim of the unfolding procedure
to determine
the true distribution $S$, given $Y$ and $M$. \\

It should be noted that $M$ only describes the migration of events. It does not 
describe losses of events, which will occur due to the finite acceptance of the 
detector, due to the trigger conditions and due to additional selections and cuts.
In an IACT experiment these losses are also determined by Monte Carlo simulations, 
and the corresponding correction factor is the effective collection area
$A_{eff}(E_{true})$. $A_{eff}(E_{true})$ has to be computed again under the same 
conditions as the experimental distribution $Y$. Apart from minor effects
(see Section \ref{section:AeffandM}), this correction can be performed quite 
independently of the unfolding. \\

The unfolding can be understood as a reshuffling of events from the bins of
$E_{est}$ into the bins of  $E_{true}$. In this procedure the
numerical values of the bin edges, both for $E_{true}$ and $E_{est}$, 
are completely irrelevant. $E_{true}$ and $E_{est}$ may be even two different
physical quantities, with completely different ranges of values and
different units, like
$E_{true}$ = (the true energy of the cosmic ray particle) and  
$E_{est}$ = (the total
number of Cherenkov photons measured for the shower) \cite{Mizobuchi2005}. 
Of course, unfolding
makes only sense if $E_{true}$ and $E_{est}$ are sufficiently
strongly correlated, otherwise the distribution of $E_{true}$ cannot be
inferred from a distribution of $E_{est}$. This is in contrast to the method
of Correction Factors (see Section \ref{section:correctionfactors}), 
where $E_{est}$ has to be a good estimate of $E_{true}$ in any case.

Because the binnings in $E_{true}$ and $E_{est}$ can be chosen
independently and to a certain degree arbitrarily (see Section \ref{section:comments}), 
the migration matrix $M$ is in general not square.  
For this reason eq. (\ref{eq:YMS}) can in general not be
inverted to obtain $S$ as $M^{-1}\cdot Y$. \\

As the unfolding is equivalent to a
reshuffling of events from the bins of the measured distribution into
the bins of the true distribution the unfolding is not restricted to
1-dimensional distributions but can in the same way be applied to
multi-dimensional distributions. The information necessary for the
unfolding procedure is completely contained in the corresponding migration
matrix, which in the case of multi-dimensional distributions
describes the migration of events from the bins of the true
multi-dimensional distribution into the bins of the measured
multi-dimensional distribution. The dimensions of the measured
and true distributions may also be different. An example for an 
unfolding in 2 dimensions is given in \cite{wittek99}.

\subsection{The direct solution of $\;\;Y\;=\;M\cdot S$}
\label{subsection:directsolution}
Very generally, the solution $S$ of the system of linear equations 
(\ref{eq:YMS}) can be obtained by minimizing the Least-Squares expression

\begin{eqnarray}
\chi_0^2\;=\;(Y-M\cdot S)^T\cdot K^{-1}\cdot (Y-M\cdot S)
\label{eq:chi20}
\end{eqnarray}
where the $nb$ components of $S$ are the free parameters. Minimizing
$\chi_0^2$ will yield solutions for $S$ which, after folding with $M$,
are best compatible with the measurement $Y$.

Two cases have to be distinguished: 
\begin{itemize}
\item {\bf The underconstrained case $nr=na\leq nb$.} \\
Because $nr=na$ the $na\times na$ matrix $G$ can be inverted and a
particular solution $S_0$ can be written as
\begin{eqnarray}
S^0\;=M^T\cdot C\qquad\qquad\qquad{\rm with}\qquad\qquad C\;=\;G^{-1}\cdot Y
\label{eq:S0}
\end{eqnarray}
If $na<nb$, the solutions $S=S_0+S_T$ form a space of 
$(nb-na)$ dimensions with $M\cdot S_T=0$. For $na=nb$ eq.(\ref{eq:S0}) 
reduces to
\begin{eqnarray}
S^0\;=\;M^T\cdot G^{-1}\cdot Y\;=\;M^{-1}\cdot Y
\label{eq:S00}
\end{eqnarray}
and $S^0$ is the only solution.
In both cases $M\cdot S=M\cdot S^0=Y$, implying $\chi_0^2=0$. The solutions are 
independent of the covariance matrix $K$. Moreover, because of (\ref{eq:normalizeM}) 
the total number of events is not changed: 
\begin{eqnarray}
\sum_i{Y_i}\;
=\;\sum_i\sum_j M_{ij}\cdot S^0_j\;=\;\sum_j{S^0_j} 
\label{eq:addconstraint}
\end{eqnarray}
\item {\bf The overconstrained case  $na> nb=nr$.} \\
Minimizing $\chi_0^2$ by varying $S$ yields
\begin{eqnarray}
S^{LSQ}\;=\;H^{-1}M^TK^{-1}\cdot Y
\label{eq:SLSQ}
\end{eqnarray}
where $H$ is the $nb\times nb$ matrix
\begin{eqnarray}
H\;=\;M^TK^{-1}M
\label{eq:H}
\end{eqnarray}
The solution $S^{LSQ}$ now depends on $K$. The minimum value of
$\chi_0^2$ becomes
\begin{eqnarray}
\chi_0^2\;=\;\left(Y-M\cdot S^{LSQ}\right)^T\cdot K^{-1}\cdot 
             \left(Y-M\cdot S^{LSQ}\right)
\label{eq:chi2LSQ}
\end{eqnarray}
Expression  (\ref{eq:SLSQ}) is also valid for $nr=na=nb$, in which case it
reduces to $S^{LSQ}\;=\;M^{-1}\cdot Y$.
\end{itemize}

It can be shown that the direct solutions (\ref{eq:S0}) and (\ref{eq:SLSQ}) may lead
to large errors of $S_j$, reflected in large absolute values of the elements of the error
matrix $T$ of $S$. This behaviour can be traced back to small eigenvalues of
the matrix $G$ and $H$ respectively \cite{anykeyev91}. \\

\section{Unfolding with Regularization - The different Unfolding Methods}
\label{section:regularization}
In order to reduce the large errors of $S$, a procedure called  regularization 
is applied.
By the regularization additional constraints are imposed on $S$, by which 
some information in the measurements $Y_i$ is discarded. 
Regularization can be viewed as a smearing of the unfolded 
distribution with some finite resolution, which reduces the correlations 
between the $S_i$ of adjacent bins at the expense that $S_i$ is no longer 
an unbiased estimate of the true distribution \cite{schmelling98}. \\

The bias increases with increasing regularization strength. Nevertheless, 
it turns out that with properly tuned 
regularization (see Section \ref{section:bestweight}) solutions can be 
obtained which are much 
closer to the true distribution than the direct solutions (\ref{eq:S0}) or
(\ref{eq:SLSQ}). \\

It is evident that regularization is particularly important in the 
underconstrained case. However, also in the overconstrained case 
regularization makes sense: even if the system of equations (\ref{eq:YMS})
is formally overconstrained ($na>nb$), it may be effectively underconstrained.
This happens for example if some of the measurements $Y_i$ have much 
larger errors than the other $Y_i$. \\

In the following, three different ways of regularization are described
\cite{anykeyev91}.

\subsection{Adding a regularization term in the expression for $\chi^2_0$}
\label{subsection:regularizationterm}
In some unfolding methods regularization is performed by adding a regularization 
term $Reg(S)$ in the expression for $\chi_0^2$ (eq.\ref{eq:chi20})
\begin{eqnarray}
\chi^2\;=\;\dfrac{w}{2}\cdot \chi_0^2\;+\;Reg(S)
\label{eq:chi2reg}
\end{eqnarray}
$w$, also called regularization parameter, is a weight which allows to steer
the regularization strength: large values of $w$ correspond to weak regularization,
small values to strong regularization. \\

\begin{itemize}
\item {\bf Tikhonov's method} \\
\label{section:tikhonov}
In Tikhonov's method \cite{tikhonov79} the regularization term is defined as
\begin{eqnarray}
Reg(S)\;=\;\sum_{j=1}^{nb}{\left(\dfrac{d^{~2}S}{dx^2}\right)_j^2}
\label{eq:regtikhonov}
\end{eqnarray}
For the second derivative $\left(\dfrac{d^{~2}S}{dx^2}\right)_j$ of
$S$ in bin $j$ different approximations may be used. The expression used
in the MAGIC software \cite{MARS} is
\begin{eqnarray}
\left(\dfrac{d^{~2}S}{dx^2}\right)_j\;
    =\;2.0\cdot\left(\dfrac{S_{j+1}-S_j}{S_{j+1}+S_j}
               -\dfrac{S_j-S_{j-1}}{S_j+S_{j-1}}\right)
\label{eq13}
\end{eqnarray}
This is actually an approximation for the bin-to-bin variation of $\Delta S/S$.
In \cite{blobel84a}  $\left(\dfrac{d^{~2}S}{dx^2}\right)_j$ is calculated from
a spline representation of $S$.

For a given value of $w$ and after specifying $Reg(S)$, expression
(\ref{eq:chi2reg}) can be minimized numerically by varying the components of
$S$. The minimization also provides the error matrix $T$ of $S$.
The regularization matrix $R$, defined in (\ref{eq13a}), can be 
calculated numerically by performing minimizations with modified
values of $Y_i$. \\

\item {\bf Schmelling's method} \\
\label{section:schmelling}
In this method, which is discussed in great detail in 
\cite{schmelling94,schmelling98}, the regularization term is set equal to the
"cross entropy"
\begin{eqnarray}
Reg(S)\;=\;\sum_{j=1}^{nb}{p_j\cdot \ln\dfrac{p_j}{\epsilon_j}}
\label{eq:regschmelling}
\end{eqnarray}
$p_j$ is the normalized distribution $S$
\begin{eqnarray}
p_j\;=\;\dfrac{S_j}{\sum_{k=1}^{nb}{S_k}}\qquad\qquad\qquad 
\sum_{j=1}^{nb}p_j\;=\;1
\label{eq14a}
\end{eqnarray}
and $\epsilon$ is a normalized prior distribution, which describes a prior
knowledge about $S$. The cross entropy $Reg(S)$ quantifies by how much $p$
deviates from $\epsilon$. Finding $S$ by
minimizing the cross entropy $Reg(S)$ simultaneously with the
least squares expression $\chi_0^2$ is called the method of 
"Reduced Cross Entropy".\\

With $Reg(S)$ from (\ref{eq:regschmelling}), expression (\ref{eq:chi2reg}) 
can now be minimized
to obtain the unfolded distribution $S$. 
Note that all components of $S$ and of the prior distribution 
are required to be $> 0$, because
otherwise $Reg(S)$ in (\ref{eq:regschmelling}) cannot be defined. 
The expressions for the error matrix $T$ of $S$ and for the regularization 
matrix $R$ are given in \cite{schmelling94,schmelling98}. 
\end{itemize}

In the MAGIC software, the condition $\sum_i{Y_i}\;=\;\sum_j{S_j}$ is used as
an additional constraint, when minimizing $\chi^2$. By this one degree 
of freedom is gained.

\subsection{Spectral Window method}
\label{subsection:spectralwindowmethod}
In some unfolding methods regularization is performed by suppresing
small eigenvalues $\lambda_l$ of $G$ by a factor $f(\lambda_l)$ 
\cite{bertero88}. By the suppression factor
$f(\lambda_l)$ the matrix $G$ and its inverse are modified. In terms of the
eigenvectors $g_l$ of $G$ they read
\begin{alignat}{2}
\widetilde{G}\;&=\;\sum_l^{nr}{f(\lambda_l)\cdot\lambda_l\cdot g_lg_l^T} \\
\widetilde{G^{-1}}\;&=\;\sum_l^{nr}{\dfrac{f(\lambda_l)}{\lambda_l}\cdot g_lg_l^T} 
\label{eq:gtilde}
\end{alignat}
where the sums extend over all eigenvalues $\lambda_l$ which are different from zero.
Like $G$ and $G^{-1}$, $\widetilde{G}$ and  $\widetilde{G^{-1}}$ are $na\times na$
matrices. 
Without suppression, $f(\lambda_l)=1$, $\widetilde{G^{-1}}$ is equal to $G^{-1}$ 
in the underconstrained case
$nr=na\leq nb$. In the overconstrained case, $na>nb=nr$, $G^{-1}$ is undefined
but $\widetilde{G^{-1}}$ can be calculated.

A similar factor $f(\kappa_l)$ can be defined to suppress small eigenvalues 
$\kappa_l$ of $H$ (eq. \ref{eq:H}). There is considerable freedom as to the choice
of the values or expressions for $f(\lambda_l)$ and $f(\kappa_l)$. 
One may introduce a parameter $i$ such that
in the limit $i\rightarrow\infty$ the suppression factors 
$f(\lambda_l, i)$ and  $f(\kappa_l, i)$ tend to 1.
$i$ has a similar meaning as the weight $w$ in eq.(\ref{eq:chi2reg}): it 
determines the regularization strength and for $i\rightarrow\infty$
the solutions tend to the direct solutions (\ref{eq:S0}) and (\ref{eq:SLSQ}) 
respectively.\\

The expressions for the error matrix $T$ of $S$ and for the regularization 
matrix $R$ are given in \cite{anykeyev91}.

\subsection{Regularization by iteration}
\label{section:bertero}
Another way of unfolding is to calculate a solution $S$
iteratively \cite{marchuk75,giljazov87,bertero88,bertero89}. 
The regularization is done by stopping the iteration at
some point. In this case the number of iterations $i$ plays a
similar role as the weight $w$ in (\ref{eq:chi2reg}). 
In the limit of an infinite number of iterations, which is equivalent 
to a very large weight $w$, the solution tends to the direct solution 
(\ref{eq:S0}) or (\ref{eq:SLSQ}) respectively. \\

\begin{itemize}
\item ${\bf nr = na\leq nb}$ \\
A simple iteration scheme \cite{anykeyev91} for solving $Y=M\cdot S$ 
with $G^{-1}\cdot Y=C$ is

\begin{eqnarray}
C^{i+1}\;=\;C^i\;-\;\tau\cdot (GC^i-Y)
\label{eq:C}
\end{eqnarray}
where $i$ is the iteration number and $\tau$ is a relaxation parameter.
The latter should be chosen in the range $0<\tau<2/\lambda_{max}$,
where $\lambda_{max}$ is the largest eigenvalue of $G$. 
Eq.(\ref{eq:C}) leads to
\begin{eqnarray}
C^i\;=\;(1-\tau G)^i\cdot C^0\;+\;
        \tau\cdot \sum_{j=0}^{i-1}(1-\tau G)^j\cdot Y\;
\label{eq22}
\end{eqnarray}
where $C^0$ is the starting value of $C$. The unfolded distribution is obtained
as $S^i\;=\;M^T\cdot C^i$. \\

In terms of the suppression factor $f(\lambda_l,i)$ this regularization can
be expressed as \cite{anykeyev91}
\begin{eqnarray}
f(\lambda_l,i)\;=\; \left[1-(1-\tau\lambda_l)^i\right]
\label{eq23}
\end{eqnarray}
if $C^0$ is set to zero, and
\begin{eqnarray}
f(\lambda_l,i)\;=\; \left[1-(1-\tau\lambda_l)^i\;+\;(1-\tau\lambda_l)^i\lambda_l\right]
\label{eq24}
\end{eqnarray}
if $C^0$ is chosen to be equal to $Y$. The solutions $S^i$ tends to
the direct solution $S^0=M^T\cdot C=M^T\cdot G^{-1}\cdot Y$ (eq. (\ref{eq:S0})). \\ 

This procedure can also be applied in the overconstrained case, $na>nb=nr$. In this
case $G^{-1}$ is undefined and it has to be replaced by $\widetilde{G^{-1}}$ 
(eq.\ref{eq:gtilde}).
The solutions $S^i$ tend to $S=M^T\cdot \widetilde{G^{-1}}\cdot Y$, with 
$f(\lambda_l)=1$. \\ 

\item ${\bf na>nb=nr}$ \\
In a similar way one may define suppression factors $f(\kappa_l,i)$ for the 
eigenvalues of the matrix $H$. In this case the solution $S$ tends to the
solution $S^{LSQ}$ (eq. (\ref{eq:SLSQ})).
\end{itemize}

In the last 2 methods, Spectral Window method and Regularization 
by iteration, 
$\chi_0^2$ doesn't appear explicitly. However, its value can be calculated and it
is taken into account when determining the optimum regularization strength 
(see Section \ref{section:bestweight}).

Once the the unfolded distribution $S_k$ is determined, by any of the methods 
described in this Section, the differential 
energy spectrum $\Phi_k$ of $\gamma$-rays is calculated 
using eq. (\ref{eq:Phik}).  $\Phi_k$ has the meaning of the average differential
$\gamma$-ray flux in the $k$-th bin of $E_{true}$ (eq.(\ref{eq:PhiE})).

\section{Forward Unfolding}
\label{section:forwardunfolding}
An implicit unfolding can be done by representing $S$ as a parametric function
$S_k(q)=f(E_k;q)$ with parameters $q=(q_1, q_2, ... q_{nq})$ and minimizing
$\chi_0^2$ in (\ref{eq:chi20})
\begin{eqnarray}
\chi_0^2\;=\;\sum_{i,j=1}^{na}
\left(Y_i-\sum_{k=1}^{nb}M_{ik}\cdot S_k(q)\right)\cdot
\left(K^{-1}\right)_{ij}\cdot   \left(Y_j-\sum_{l=1}^{nb}M_{jl}\cdot
                                                          S_l(q)\right)
\label{eq6a}
\end{eqnarray}
with respect to the parameters $q$. The number of measurements is equal to $na$,
the number of unknowns $nq$. Thus, the problem is overconstrained if
$nq<na$, independent of the value of $nb$. One degree of freedom is gained if the
total number of events is required to stay constant: $\sum_j S_j\;=\;\sum_i Y_i$.

In many cases the
minimization of $\chi_0^2$ can be performed analytically, by solving
$\dfrac{\partial \chi_0^2}{\partial q}=0$, similarly to the procedure
described in Section \ref{subsection:directsolution}.

The parametrization of $S$ can be written in the form
\begin{eqnarray}
S_k(q)\;=\;\Phi_k(q)\cdot T_{eff}\cdot \Delta(E_{true}^k)
\cdot A_{eff}^k\cdot A_{addcut}^k\cdot A_{absorption}^k 
\label{eq:Phik}
\end{eqnarray}
with
\begin{eqnarray}
\Phi_k(q)\;&=\;\dfrac{\int_{\Delta(E_{true}^k)}\Phi(E_{true},q)\cdot dE_{true}}
                  {\Delta(E_{true}^k)} 
\label{eq:PhiE}
\end{eqnarray}
Here $\Phi(E_{true},q)$ is the assumed parametrization of the differential 
energy spectrum of $\gamma$-rays, 
$T_{eff}$ is the effective observation time, $\Delta(E_{true}^k)$ denotes
the $k$-th bin in $E_{true}$ or its width, 
$A_{eff}^k$ is the effectice collection area,
and $A_{addcut}^k$ is the reduction 
factor due to an additional cut (for example in $E_{est}$, see Section
\ref{section:additionalcuts}). A further
correction  $A_{absorption}^k$ can be introduced, if $\Phi(E_{true},q)$ is
supposed to represent the differential energy spectrum of $\gamma$-rays before
absorption, either at the $\gamma$-source or by interaction with the extragalactic
photon background. $A^k_{addcut}$ can be determined from MC simulations, whereas
$A^k_{absorption}$ can be calculated in models about the extragalactic 
photon background \cite{kneiske04}. 

Parametrizing $S$ as an analytic function of $E_{true}$, with some 
free parameters $q$, can be understood as a kind
of regularization, because it forces the solution $S$ and its
derivatives to be continuous, leading to a suppression of the noise
component of $S$. 

The Forward Unfolding does not provide an unfolded distribution $S$. 
It provides those parameter values $q$ for an assumed parametrization of 
$\Phi(E_{true})$, which minimize  $\chi_0^2$  in (\ref{eq6a}). Of course,
$S$ can then be calculated from $\Phi(E_{true})$ via (\ref{eq:Phik}).

Under the assumption of a certain parametrization of $\Phi(E_{true})$,
the Forward Unfolding is a very robust method of determining the best 
parameter values $q$. Moreover, since there is no regularization strength
to be adjusted, the uncertainty as to its choice does not exist.
Therefore Forward Unfolding represents a powerful and useful check
of the unfolding results obtained by any of the methods described in Section
\ref{section:regularization}. In those methods the parametrization is only 
introduced after the actual unfolding of the measurements $Y$.  

\section{Useful quantities in the unfolding}
\label{section:usefulquantities}
In this Section two quantities are explained which are useful for judging
the quality of the unfolding result: The error matrix (covariance matrix) 
$T$ of the unfolded 
distribution $S$ and the regularization matrix $R$. 

\subsection{The covariance matrix of the unfolded distribution $S$}
In all unfolding methods the covariance matrix $T$ of $S$ can be determined.
The trace of $T$, $Trace(T)$, measures the noise component of $S$,
as $Trace(K)$ measures the noise component of $Y$.

In the methods where the solution $S$ is given as 
$S\;=\;D\cdot Y$, like in Schmelling's method or in the Spectral Window method,
the covariance matrix $T$ of $S$ is obtained by
$T\;=\;D\cdot K\cdot D^T$. \\

In those methods where the solution is determined 
by a numerical minimization of $\chi^2$, like in Tikhonov's method,
$T$ is obtained from the shape of $\chi^2$ in the region around the minimum.

\subsection{The regularization matrix $R$}
\label{section:regularizationmatrix}
A quantity which describes how the estimates  $\sum_{k=1}^{nb}M_{jk}S_k$ 
of $Y_j$ couple to the
measurements $Y_i$ is given by the $na\times na$ matrix
\begin{eqnarray}
R_{ij}\;=\;\dfrac{\partial\left( \sum_{k=1}^{nb}M_{jk}S_k\right)}
                     {\partial Y_i}\qquad\qquad{\rm or}\qquad\qquad
R\;=\;\dfrac{\partial(M\cdot S)}{\partial Y}
\label{eq13a}
\end{eqnarray}
also called regularization matrix \cite{schmelling98}. The trace of $R$
can be interpreted as the effective number of measurements used in the 
unfolding procedure. The number of effectively rejected measurements is then
equal to $N_{rej}=na-Trace(R)$.\\

The maximum value of $Trace(R)$, which is equal to the rank $nr$ of $G$, 
is reached with the direct solutions  
(\ref{eq:S0}) and (\ref{eq:SLSQ}), corresponding to the cases $nr=na\leq nb$ and
$na>nb=nr$ respectively:
For the direct solution $S^0$ the number of rejected measurements
$N_{rej}$ is equal to $na-Trace(R)\;=\;na-nr\;=\;na-na\;=\;0$, which means that no 
information is discarded. The
measurements $Y$ are completely reproduced by the unfolding :
$M\cdot S^0\;=\;Y$.
For the least squares solution $S^{LSQ}$ the number of effectively
rejected measurements is $N_{rej}\;=\;na-Trace(R)\;=\;na-nr\;=\;na-nb\;>\;0$, which means that
some information is discarded. The
measurements $Y$ are not exactly reproduced by the unfolding :
$M\cdot S^{LSQ}\;\neq\;Y$. The fact that the system is overconstrained
has a similar effect as regularization. 
In both cases, with increasing regularization strength $Trace(R)$ is reduced 
and $N_{rej}$ is increased.

\section{Selecting the unfolding result}
\label{section:bestweight}
For a given unfolding method the result $S$ depends on the
regularization strength, which is given by the
weight $w$ (or by the number of iterations $i$ respectively). In the
literature various criteria for choosing the "best" weight are
proposed \cite{belogorlov85,zhigunov88,anykeyev91,schmelling98,blobel02,
hoecker96}. 
Unfortunately, none of them provides a choice which is optimal for all cases.
Reasons for this are: The optimum regularization strength 
in general depends
on the shape of the unknown distribution $S$. It also depends on the binnings in
$E_{true}$ and $E_{est}$ and on the prior distribution (if applicable). \\

\begin{figure}[h!]
\centering
\subfigure[$\chi_0^2$] {\label{1a}
                        \includegraphics[width=0.48\textwidth]{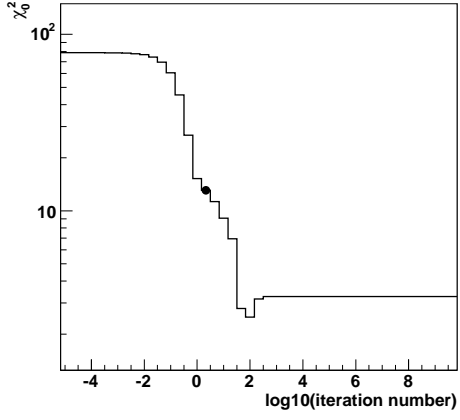}}
\subfigure[$Trace(T)/Trace(K)$] {\label{1b}
                        \includegraphics[width=0.48\textwidth]{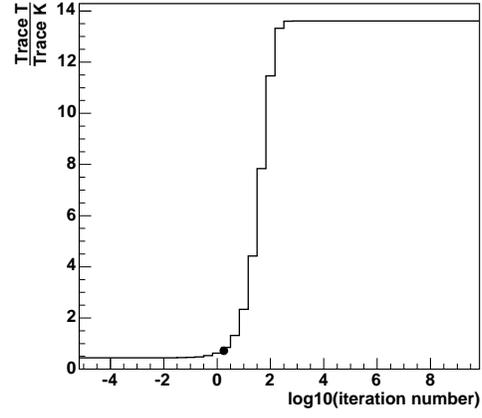}}
\subfigure[$Trace(R)$] {\label{1c}
                        \includegraphics[width=0.48\textwidth]{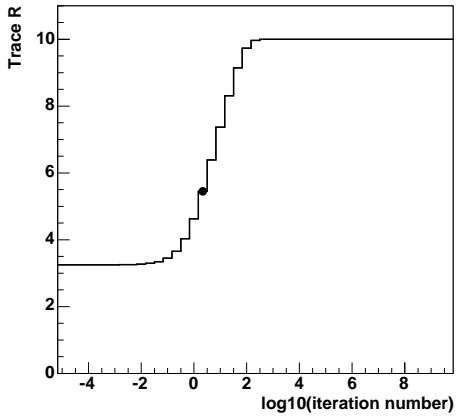}}
\subfigure[$Reg(S)_{Tikhonov}$] {\label{1d}
                        \includegraphics[width=0.48\textwidth]{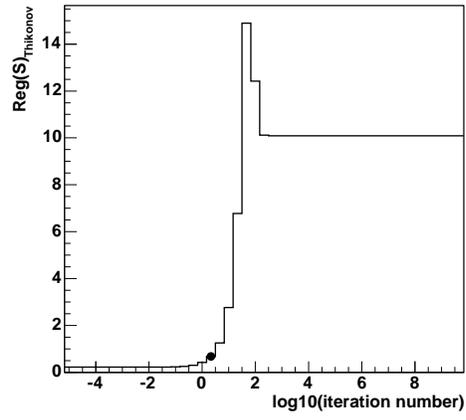}}
\subfigure[$Reg(S)_{Schmelling}$] {\label{1e}
                        \includegraphics[width=0.48\textwidth]{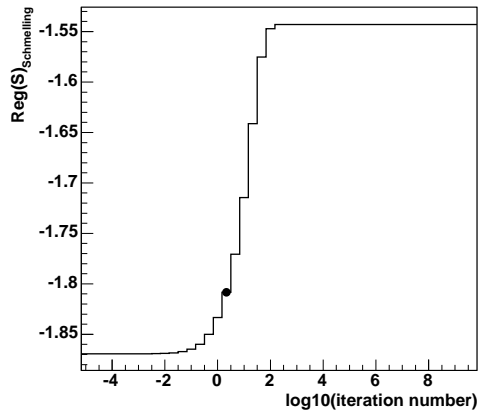}}
\caption[XXX]{Useful quantities for determining the optimum regularization
strength, plotted as a function of the iteration number $i$.}
\label{fig:iterbertero}
\end{figure}

The effect of the regularization is illustrated in
Fig. \ref{fig:iterbertero}, where
different quantities are plotted as a function of the iteration number $i$. 
In this example an experimental energy distribution of $\gamma$-rays from the
Crab Nebula \cite{albert07}, i.e. the number of excess events in bins of the 
estimated energy
$E_{est}$, was unfolded. The unfolding was performed for 30
different $i$ in the range $10^{-5}$ to $10^{10}$, using the method of
Bertero (eq.(\ref{eq24})). More results from the analysis of these data
are given in Section \ref{section:application}.
With decreasing $i$, i.e with increasing 
regularization strength, one observes an increase of $\chi_0^2$ 
(eq. \ref{eq:chi20}) and a decrease 
in the quantities $Trace(T)/Trace(K)$, $Trace(R)$, $Reg(S)_{Tikhonov}$ 
(eq. \ref{eq:regtikhonov}) and $Reg(S)_{Schmelling}$ (eq. \ref{eq:regschmelling}).
Very similar behavior is found for the other unfolding methods, discussed in Section
\ref{section:regularization}.

Obviously, an acceptable unfolding result should satisfy the following 
conditions:
\begin{itemize}
\item The $\chi^2$-probability, calculated from  the value of
$\chi_0^2$ and the number of degrees of freedom in the unfolding procedure,
should be acceptable, say $>1\%$. Otherwise
the unfolding result is incompatible with the measured distribution
$Y$. \\
\item The noise term $Trace(T)$ of the unfolded distribution $S$ should be 
comparable to the noise term $Trace(K)$ of the measurements. The main 
aim of regularization is a suppression of the large noise term of $S$, 
which one often obtains if no regularization is applied.
A large noise term $Trace(T)$, as well as large correlation terms of $T$, 
indicate a too fine binning in $E_{true}$, leading to small eigenvalues 
of $G$ or $H$ (see Section \ref{sect:optimalbinnings}). \\
\item $Trace(R)$ should not be much lower than its maximum possible
value, which is equal to the rank $nr$ of the matrix $G$.  Otherwise the 
solution is too strongly dominated and biased by the regularization.
\end{itemize}

For determining the "best" regularization strength a compromise has to 
be found between the above requirements. It has turned out that the 
criterion $Trace(T)\;=\;Trace(K)$ in general leads to 
solutions which satisfy the above conditions reasonably well, provided 
the problem is not strongly overconstrained. In the latter case, where the 
unfolding result is better constrained, a solution with $Trace(T)\;<\;Trace(K)$
is more apropriate. In MAGIC the standard criterion for determining the 
optimal regularization strength is  $Trace(T)\;=\;Trace(K)$. The full circles 
in Fig. \ref{fig:iterbertero} indicate this choice.
However,
any other regularization strength can be chosen by hand, if this is suggested
by the behaviour of the quantities $\chi^2_0,\;Trace(T),\;Trace(R),
\;Reg(S)_{Tikhonov}$ or $\;Reg(S)_{Schmelling}$.

Unfolding with regularization is a procedure which allows freedom in the 
choice of the regularization method and in the choice of the regularization 
strength. The above criteria for an acceptable solution strongly restrict 
this freedom. Nevertheless a certain degree of arbitrariness remains as to 
which unfolding result should be considered representative and final.
In MAGIC a selected unfolding result is considered representative if all other 
unfolding methods yield results, which are also acceptable and statistically 
consistent with the selected result. In addition, it is required that also the 
Forward Unfolding (Section \ref{section:forwardunfolding}), 
using a reasonable parametrization of $\Phi(E_{true})$,
gives a consistent result. 

An uncertainty due to the unfolding
is determined from the spread of the $S_j$, obtained from the different unfolding 
methods.

\section{Further comments on the unfolding}
\label{section:comments}
In the actual application of the unfolding procedure to real data some technical
details have to be considered, and they are discussed in this Section.

\subsection{Optimal binnings}
\label{sect:optimalbinnings}
The binning of the experimental distribution $Y$ is often dictated by
the available statistics and by the experimental errors. The binning
should not be chosen too fine in order to assure significant
measurements in all bins. In the case of an IACT experiment, a 
sufficiently large sample
is required to determine the number of signal (excess)
events with sufficient accuracy. The binning should not be chosen too
wide either, because the binning in $Y$ limits the reconstruction of
the fine structure of the unfolded distribution $S$.

Another criterion for the binnings is the behavior of Gram's matrix
$G$. A too fine binning for $S$ leads to strong
correlations between neighboring columns of the migration
matrix $M$, implying small eigenvalues of $G$, which lead to a large
noise component  of $S$. Given a certain choice of the 
binning in $E_{est}$ and thus of $na$, the bin size in $E_{true}$ 
or $nb$ should be set such, that the system of linear equations
(\ref{eq:YMS}) is not underconstrained. This usually leads to wider 
bins in $E_{true}$ than in $E_{est}$. In MAGIC a typical value of 
$\Delta log_{10}(E_{true})/\Delta log_{10}(E_{est})$ is 1.4 (see Section
\ref{section:application}). \\

It should also be noted that the unfolding procedure doesn't require 
equidistant bins, neither in $E_{est}$ nor in $E_{true}$.

\subsection{Completeness of the migration matrix}
\label{sect:completeness}
If $na1\leq i \leq na2$ specifies the range of bins of the measured
distribution $Y$ which are to be considered in the unfolding procedure, also 
the range in $E_{true}$ to be considered in the unfolding has to be
chosen properly: one has to make sure that all bins $j$ of $E_{true}$
are present, for which the column $j$ of the 
migration matrix $M_{ij}$ contributes to the selected bins of $Y$, 
i.e. for which at least one of
the elements $M_{ij}\;\;(na1\leq i \leq na2)$ is different from zero. 

There is an exception to this rule, if for some reason certain bins $j$ 
of $E_{true}$ are not expected to contribute to the selected bins in $E_{est}$.
This is for example the case if one of the factors $A_{xxx}^k$ in 
(\ref{eq:Phik}) is so small that $S_k$ can be neglected.

\subsection{Additional cuts}
\label{section:additionalcuts}
As explained in Section \ref{section:notation}, the distribution
$Y$ and the migration matrix $M$ have to be produced under identical
conditions (selections, cuts). If an additional cut is imposed when
generating $Y$, also $M$ has to be recalculated before doing the
unfolding. If this additional cut is a cut in $E_{est}$ one may 
proceed in the following way:
\begin{itemize}
\item Renormalize the columns $j$  of $M$ to the selected range in $E_{est}$
(see eq.(\ref{eq:normalizeM})).
\item Perform the unfolding of $Y$ in the usual way.
\item Apply a correction $A_{addcut}^j$ to the unfolded 
distribution $S_j$, where  $A_{addcut}^j$ is the renormalization 
factor for column $j$ of $M$.
\end{itemize}

\subsection{Starting values for the minimization}
In the cases where the unfolding procedure involves numerical
minimizations, like the ones discussed in section
\ref{subsection:regularizationterm}, the minimization
may not converge. This
problem can be often solved by choosing different starting
values. Another reason for non-convergence is discussed in Section 
\ref{section:cannotbedetermined}. In MAGIC the standard choice of the 
starting distribution for $S$, and also of the prior distribution
$\epsilon$ in Schmelling's method, 
is a distribution which is close to the measured distribution $Y$.

\subsection{Components of $S$ which cannot be determined in the
unfolding}
\label{section:cannotbedetermined}
According to eq. (\ref{eq:YMS}) those $S_j$ for which the column $M_{ij}$
$(i=1, ...na)$
is a null vector have no influence on $Y$ and can therefore not be determined in
the unfolding. These components should not be varied in the minimization
because they would unnecessarily complicate the minimization process
and may lead to non-convergence.

\subsection{Dependence on the assumptions made in the Monte
Carlo simulation}
\label{section:MCdependence}
For the unfolding the migration matrix $M_{ij}$ is the crucial quantity.
Obviously, if it doesn't describe the real migration of events
correctly, the unfolding result will be wrong. This means that at
fixed $j$, i.e. at fixed $E_{true}$, the MC simulation has to 
describe the migration in $E$ correctly. This will be the case if at
fixed $E_{true}$ the shower simulation is realistic and if the detector
response is simulated correctly.

On the other hand, the
distribution of $E_{true}$ in the MC need not agree with the real
distribution of $E$ : due to the normalization of $M$ 
(eq. (\ref{eq:normalizeM})) the bin-to-bin-variation of the number of 
MC events in $E_{true}$ has no
influence on $M_{ij}$ at all. This is one of the great advantages of
unfolding methods like those presented in Section \ref{section:regularization} 
as compared to methods based on correction factors (see Section 
\ref{section:correctionfactors}).

However, there is a residual dependence of $M_{ij}$ due to the finite binning
in $E_{true}$ : Depending on the shape of the $E_{true}$ distribution
within an $E_{true}$ bin in the MC simulation, the calculated $M_{ij}$ may be more
representative for the lower, middle or upper part of the $E_{true}$ bin. 
If the $E_{true}$ distribution in the real data is different from that
in the MC simulation the calculated $M_{ij}$ may not be exactly the
right ones. 

This residual dependence of $M$ on the shape of the $E_{true}$
distribution in the MC can be nearly completely removed by an
iteration procedure in which the $M$ for the next iteration step is
determined from a MC sample, in which the $E_{true}$ distribution has
been corrected using the unfolding result of the last iteration step
(see Section \ref{section:AeffandM}).

\section{Determining the effective collection area $A$ and the
migration matrix $M$ for a finite bin in $E_{true}$ and $\Theta$}
\label{section:AeffandM}
In an IACT experiment it is important that the effective collection 
area $A$, which enters in the flux calculation, is computed taking into
account a realistic shape of the differential flux $\Phi(E_{true})$ 
and the actual distribution of
effective observation times $dT(\Theta)/d\Theta$ in the zenith angle
$\Theta$. This is also important for the migration matrix $M$, which enters
in the unfolding, because of the residual dependence on the flux
spectrum, as discussed in Section
\ref{section:MCdependence}. Recalculating $A$ and $M$ with the proper
$E_{true}$ and $\Theta$ spectra is the more important the bigger the
$(\Delta E_{true},\;\Delta\Theta)$ interval and the stronger the
variations of $M$ and $A$ within this bin are. Often, for statistics reasons,
large bin sizes in $E_{est}$ and $\Theta$, and thus also in $E_{true}$,
have to be chosen.  

In the following it is assumed that the effective collection area $A$
and the migration matrix $M$ 
are known functions of $E_{true}$ and $\Theta$. This can be
achieved by determining them from a sample of MC $\gamma$-ray events 
in very fine bins of $E_{true}$
and $\Theta$. The aim is to calculate an average $\overline{A}$ of the
effective collection area and an average $\overline{M}$ of the
migration matrix, which are representative for a finite bin 
$(\Delta E_{true},\;\Delta\Theta)$ in $E_{true}$ and $\Theta$.

\subsection{\bf The effective collection area} 
The number of observed events in a $(\Delta E_{true},\;\Delta\Theta)$ 
bin is given by
\begin{alignat}{2}
S\;=\;\int_{\Delta\Theta}\int_{\Delta E_{true}}
               A(E_{true},\Theta)\cdot\Phi (E_{true}) 
               \cdot\dfrac{dT(\Theta)}{d\Theta}
               \cdot dE_{true}\cdot d\Theta
\label{eq41}
\end{alignat}
Here $\Phi (E_{true})$ is the differential $\gamma$-ray flux to be
measured, $\dfrac{dT(\Theta)}{d\Theta}$ is the distribution of
observation times in the experimental data and $A(E_{true}, \Theta)$
is the known dependence of the effective collection area on $E_{true}$ and
$\Theta$. 

With the definitions
\begin{alignat}{2}
\Delta T\;&=\;\int_{\Delta\Theta}{\dfrac{dT(\Theta)}{d\Theta}\cdot d\Theta} 
\qquad\qquad ({\rm =\;total\;observation\;time}) \label{eq42} \\
\nonumber \\
\overline{\Phi}\;&=\;\dfrac{1}{\Delta E_{true}}\;\;\int_{\Delta E_{true}}
\Phi(E_{true})\cdot dE_{true}   \label{eq42a}\\
\nonumber \\
\overline{A}\;
              &=\;\dfrac{\int_{\Delta\Theta}\int_{\Delta E_{true}}
               A(E_{true},\Theta)\cdot\Phi (E_{true}) 
               \cdot\dfrac{dT(\Theta)}{d\Theta}
               \cdot dE_{true}\cdot d\Theta}
            {\Delta T\cdot\overline{\Phi}\cdot\Delta E_{true}} 
\label{eq44x} 
\end{alignat}
equation (\ref{eq41}) can be rewritten as
\begin{alignat}{2}
\overline{\Phi}\;=\;\dfrac{S}{\Delta T\cdot \overline{A}\cdot \Delta E_{true}}
\label{eq47}
\end{alignat}
This is the usual formula for converting numbers of events $S$ into differential fluxes
$\Phi$. Because of the definitions (\ref{eq42a}) and (\ref{eq44x}) the differential
flux $\overline{\Phi}$ in an $E_{true}$ bin, as determined in the unfolding,
is the average differential flux in this bin. Therefore, when quoting or plotting
a result for $\overline{\Phi}$ the bin edges in $E_{true}$ should also be given or shown.

\subsection{\bf The migration matrix}
The number of reconstructed MC events in a bin $i$ of $E_{est}$ can be written as
\begin{alignat}{2}
N_i\;=\;\int_{\Delta\Theta}\int_{\Delta E_{true}}
M_i(E_{true},\Theta)\cdot &A(E_{true},\Theta)
\cdot \Phi (E_{true})\cdot\dfrac{dT(\Theta)}{d\Theta}
\cdot dE_{true}\cdot d\Theta \label{eq59}\\
{\rm with}\qquad  \sum_kM_k(E_{true},\Theta)\;&=\;1\qquad\qquad{\rm for\;all\;}
E_{true}\;{\rm and\;}\Theta
\nonumber
\end{alignat}

$M_i(E_{true},\Theta)$ is the element of the normalized migration matrix
for the $i$-th bin in $E_{est}$ , at an energy $E_{true}$. The index $j$ of
$M_{ij}$ is replaced by the variable $E_{true}$.
The dependence of $M_i(E_{true},\Theta)$ on $E_{true}$ and $\Theta$ is assumed 
to be known from MC simulations. 

The average migration matrix $\overline{M}_i$ for the selected $\Delta E_{true}$
bin, to be used in the unfolding, is then obtained by  
\begin{alignat}{2}
\overline{M}_i\;&=\; \dfrac{N_i}{\sum_k N_k} \nonumber \\
              \;&=\;\dfrac
{\int_{\Delta\Theta}\int_{\Delta E_{true}} M_i(E_{true},\Theta)\cdot 
A(E_{true},\Theta)
\cdot \Phi (E_{true})\cdot\dfrac{dT(\Theta)}{d\Theta}
\cdot dE_{true}\cdot d\Theta}
{\int_{\Delta\Theta}\int_{\Delta E_{true}}
A(E_{true},\Theta)
\cdot \Phi (E_{true})\cdot\dfrac{dT(\Theta)}{d\Theta}
\cdot dE_{true}\cdot d\Theta}
\label{eq60}
\end{alignat}

The averages  $\overline{M}_i$ and  $\overline{A}$ 
are calculated according to (\ref{eq60}) and (\ref{eq44x}) respectively, 
using an approximation $\Phi_1 (E_{true})$ of the function $\Phi (E_{true})$.
The measured distribution $Y$ is unfolded using $\overline{M}_i$, yielding 
the unfolded distribution $S$. 
A new approximation  $\Phi_2 (E_{true})$ is then determined from $S$ according to 
eq. (\ref{eq47}). The procedure is iterated until  
$\Phi (E_{true})$ has converged. In practice parametric functions are used as
approximations of $\Phi (E_{true})$, and in general the convergence 
is found to be very fast.

\section{Combining data before applying the unfolding procedure}
\label{section:combine}
One often has the situation that there exist several measured
distributions $Y^{\nu}$ of the same quantity. If the $Y^{\nu}$ were
obtained under different conditions also the migration matrices
$M^{\nu}$ and the effective collection areas $A^{\nu}$
will be different for the different measurements. 

In an IACT experiment the different conditions may be
\begin{itemize}
\item Different modes of observation (ON/OFF mode, wobble mode, observation in the presence of moon light,
...).
\item Different ranges of the zenith angle.
\item Different detector conditions.
\item etc.
\end{itemize}

In order to determine a final unfolded distribution $S$ one may proceed in
different ways :
\begin{itemize}
\item {\bf Individual unfolding} : \\
Unfold each $Y^{\nu}$ using $M^{\nu}$ to obtain $S^{\nu}$, and combine
the $S^{\nu}$ to obtain the final solution $S$. \\

\item {\bf Global unfolding} : \\
Combine the $Y^{\nu}$ and $M^{\nu}$ to obtain a global $Y$ and $M$, do
an unfolding of $Y$ using $M$, which will give the final solution $S$.
\end{itemize}
There is one important argument in favour of the second option : Each
of the $Y^{\nu}$, or some of them, may have large statistical errors,
making the unfolding of the individual $Y^{\nu}$ unstable. One common
unfolding of the global $Y$ using the global $M$ will be more robust,
in general.

In the case of an IACT experiment, the following relations hold for
each measurement $\nu$ :
\begin{alignat}{2}
  Y^{\nu}\;  &=\; M^{\nu}\cdot S^{\nu}  \nonumber \\
  S^{\nu}_j\;&=\; T^{\nu}\cdot A^{\nu}_j\cdot \Phi_j\cdot \Delta E_j
\label{eq31}
\end{alignat}
$j$ denotes the $j$-th bin in $E_{true}$, $\Delta E_j$ is its width,
$\Phi_j$ is the average flux 
and $A_j$ is the effective collection area in
this bin, and $T$ is the effective observation time. Since $\Phi$ and
$\Delta E$ are the same for all $\nu$ the relations
for the combined data read :
\begin{alignat}{2}
  Y\;  &=\; M\cdot S  \nonumber \\
  S_j\;&=\; T\cdot A_j\cdot \Phi_j\cdot \Delta E_j
\label{eq32}
\end{alignat}
Inserting $Y=\sum_{\nu}{Y^{\nu}}$ and  $T\;=\;\sum_{\nu}T^{\nu}$
in (\ref{eq32}) and using (\ref{eq31}) 
one obtains
\begin{alignat}{2}
  A_j\;   &=\;\dfrac{\sum_{\nu}{A^{\nu}_j\cdot T^{\nu}}}{T} \label{eq32a} \\
  M_{ij}\;&=\;\dfrac{\sum_{\nu}{M^{\nu}_{ij}\cdot(A^{\nu}_jT^{\nu})}}
                    {\sum_{\mu}{(A^{\mu}_jT^{\mu})}}
        \;=\;\dfrac{\sum_{\nu}{M^{\nu}_{ij}\cdot(A^{\nu}_jT^{\nu})}}
                   {A_jT}  
\label{eq33}
\end{alignat}
The relations (\ref{eq32a}) and (\ref{eq33}) give the prescription how
to combine the
individual $M^{\nu}$ to obtain the global $M$, and how to combine the
individual $A_j^{\nu}$ to obtain the global $A_j$. The measured distribution
$Y$ is unfolded using $M$, and the unfolded distribution $S$ is converted
into a flux $\Phi$ according to (\ref{eq32}).
As can be seen from (\ref{eq32a}) and (\ref{eq33}) the $A_j$ are
weighted averages of the $A_j^{\nu}$ with weights
$w_{\nu}\;=\;T^{\nu}/T$, and the $M_{ij}$ are weighted averages of the
$M_{ij}^{\nu}$ with weights
$w_{\nu}\;=\;A_j^{\nu}T^{\nu}\;/\;(\sum_{\mu}A_j^{\mu}T^{\mu})$. The
weights can be interpreted as fractions because they add up to 1. They
are correlated and their covariance matrix has to be taken into
account when calculating the errors of $A_j$ and $M_{ij}$.

The equations (\ref{eq32a}) and (\ref{eq33}) also show that $A_j$ and
$M_{ij}$ can be obtained without knowning which spectra
$\Phi(E_{true})$ and $dT(\Theta)/d\Theta$ were used to compute the
$A^{\nu}_j$ and $M^{\nu}_{ij}$.

\section{Application to experimental data}
\label{section:application}
\begin{figure}[h!]
\centering
\subfigure[Original migration matrix $M$] {\label{2a}
              \includegraphics[width=0.48\textwidth]{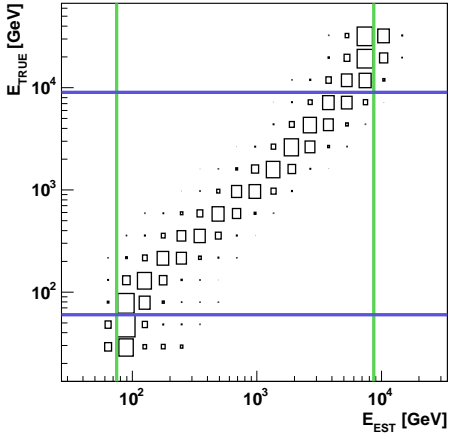}}
\subfigure[Distribution $Y$ to be unfolded (open circles), and the unfolded distribution
$S$ folded with $M$ (histogram)] {\label{2b}
              \includegraphics[width=0.48\textwidth]{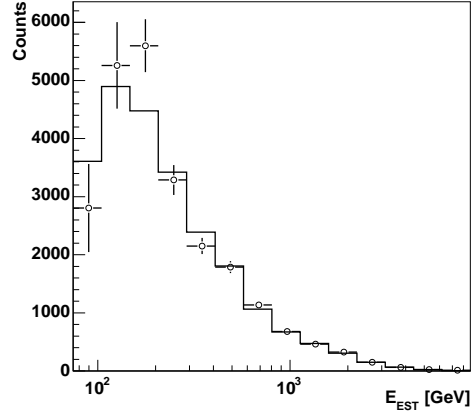}}
\subfigure[Effective collection area $A_{eff}$] {\label{2c}
              \includegraphics[width=0.48\textwidth]{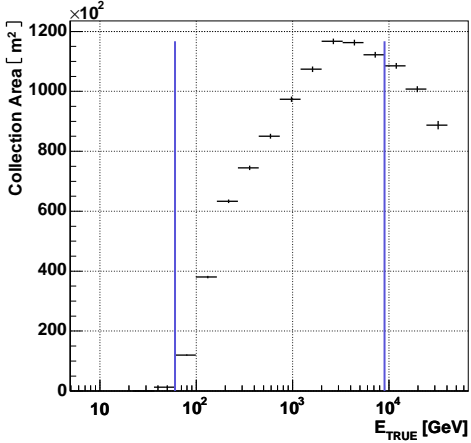}}
\subfigure[Eigenvalues of Gram's matrix $G$] {\label{2d}
              \includegraphics[width=0.48\textwidth]{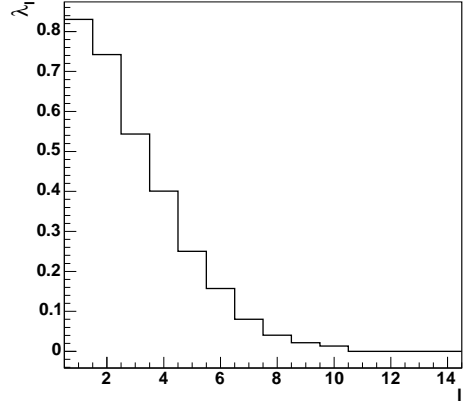}}
\caption[XXX]{Input data for the Unfolding and for the calculation of the differential
$\gamma$-ray flux.}
\label{fig:CrabData}
\end{figure}

In this Section an experimental energy distribution of $\gamma$-rays from the
Crab Nebula is unfolded, which was obtained in an analysis of data taken with 
the MAGIC telescope \cite{albert07}. The migration matrix $M$, as determined from 
a sample of $\gamma$-MC events, is plotted in Fig. \ref{fig:CrabData}a) as a function 
of $E_{est}$ and $E_{true}$. The size of the boxes is proportional to the value
of $M_{ij}$, where $i$ and $j$ are the bin numbers in $E_{est}$ and $E_{true}$
respectively. The experimental distribution $Y$ of the number of $\gamma$-excess
events as a function of $E_{est}$ is displayed in Fig. \ref{fig:CrabData}b). Both 
distributions are after all cuts and selections, except a cut in $E_{est}$.

\begin{figure}[h!]
\subfigure[Unfolded distribution $S$ before (grey symbols) and after
           the correction for the cut in $E_{est}$ (black symbols).] {\label{3a}
              \includegraphics[width=0.48\textwidth]{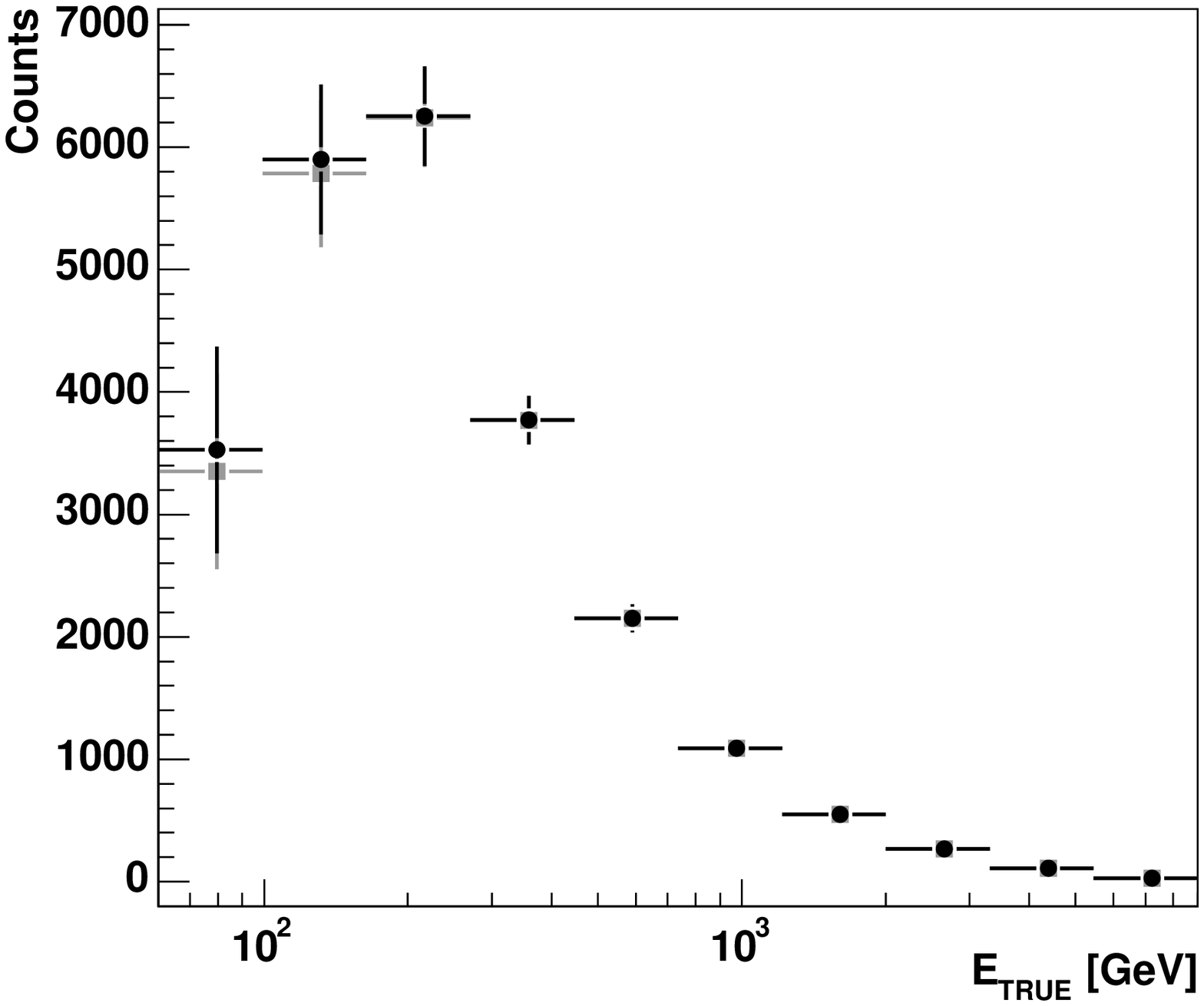}}
\subfigure[Result for the differential $\gamma$-ray flux $\Phi (E_{true})$ 
           multiplied with $E^2$.] {\label{3b}
              \includegraphics[width=0.48\textwidth]{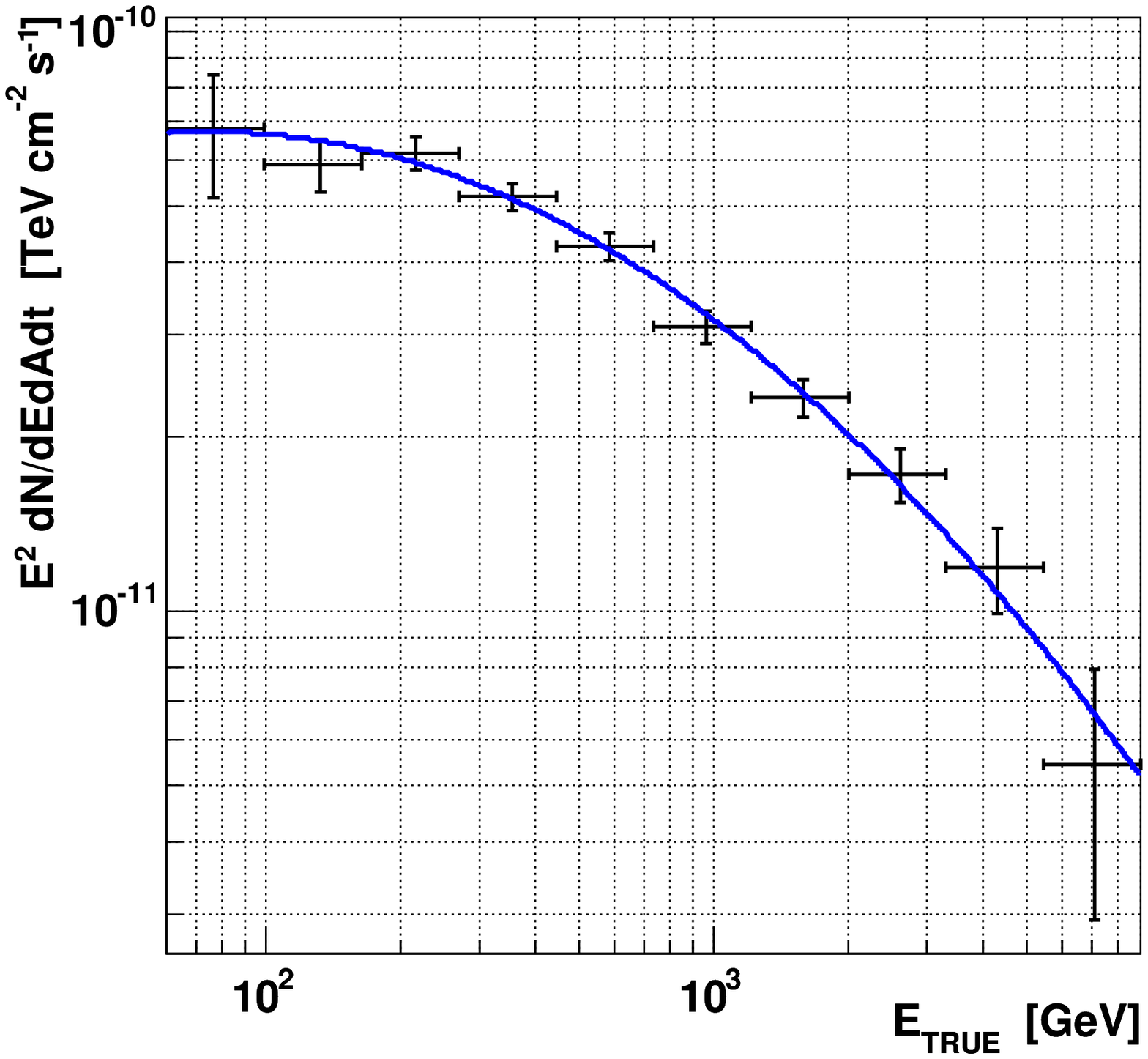}}
\subfigure[Comparison of the results for $E^2\cdot\Phi (E_{true})$
           from different unfolding methods.] {\label{3c}
              \includegraphics[width=0.48\textwidth]{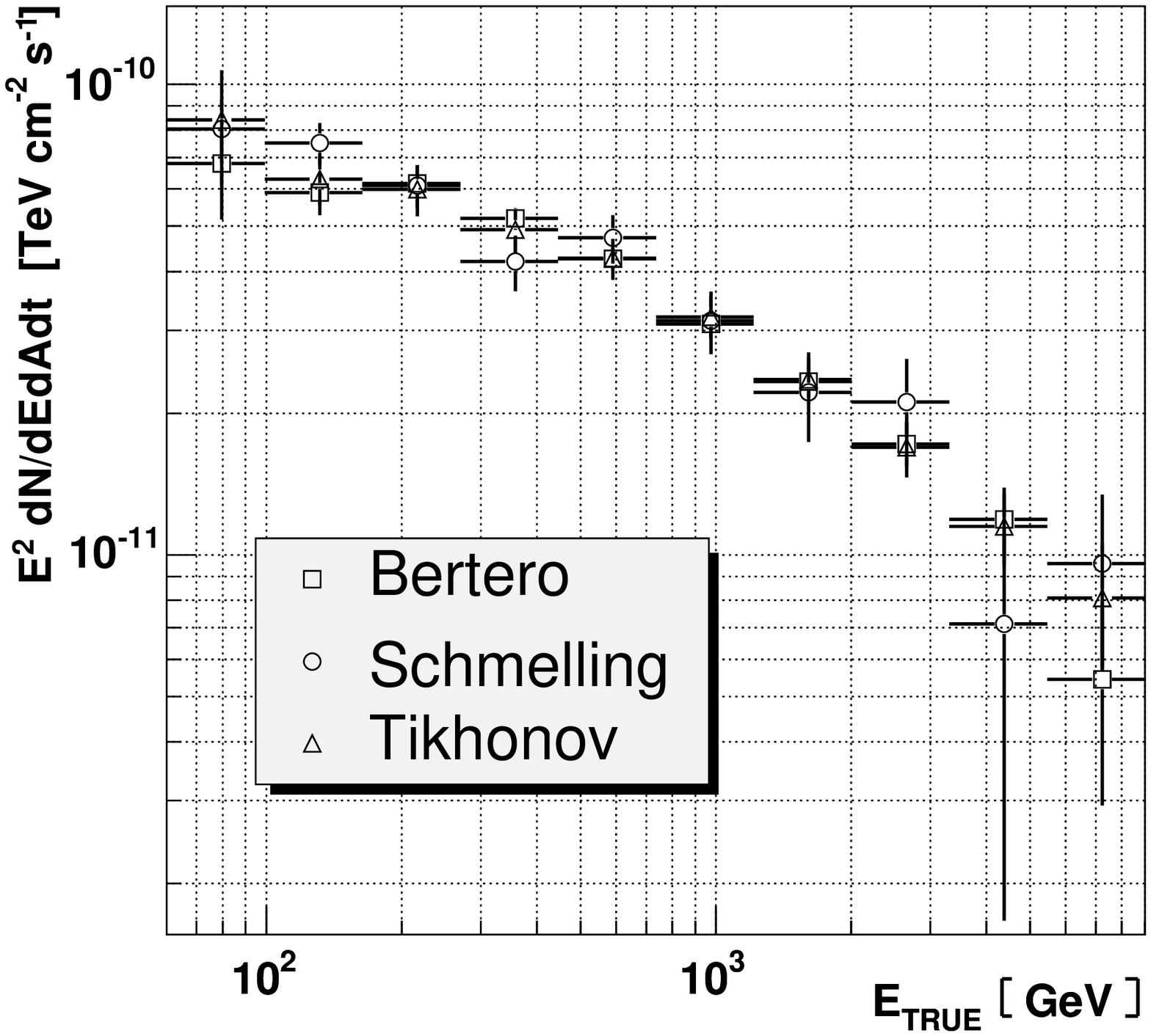}}
\caption[XXX]{Results from the Unfolding.}
\label{fig:CrabResults}
\end{figure}

The vertical and horizonthal lines in Figs. \ref{fig:CrabData}a) and c) indicate
the ranges in $E_{est}$ or $E_{true}$, 
which were selected for the unfolding.
The range in $E_{est}$ is given by those $E_{est}$ bins for which a significant 
number of excess events could be determined. The range in $E_{true}$ comprises 
all those $E_{true}$ bins which are expected to contribute to the selected
range in $E_{est}$. Fig. \ref{fig:CrabData}a) suggests that these are all 15 $E_{true}$
bins. However, according to the plot of the effective collection area $A_{eff}$ 
in Fig. \ref{fig:CrabData}c) the contribution from $E_{true}<60$ GeV is expected to be 
negligible. The same holds for $E_{true}>9$ TeV, due to the strongly decreasing 
$\gamma$-ray flux with increasing energy. This is confirmed by a Forward Unfolding
of $Y$ in which a differential $\gamma$-ray flux of the form 
$\dfrac{dN}{dA\cdot dt\cdot dE}=f_0\cdot \left(\dfrac{E}{300\;GeV}\right)^{\alpha}$, 
with $\alpha=a+b\cdot\log_{10}\left(\dfrac{E}{300\;GeV}\right)$,
was assumed. This leads
to a number of $E_{est}$ and $E_{true}$ bins of $na=14$ and $nb=10$ respectively,
which are used in the unfolding procedure. The size of the $log_{10}(E_{true})$ bins was 
deliberately chosen wider than the $log_{10}(E_{est})$ bins by a factor of 1.4,
in order to better constrain the unfolding. The rank of Gram's matrix $G$ is 
equal to $nr=10$, as can be seen from Fig. \ref{fig:CrabData}d), which shows the size
of the eigenvalues $\lambda_l$ of $G$ as a function of $l$. Two to three of these
eigenvalues are much smaller than the maximum eigenvalue, and they are the 
reason for the large values of $Trace(T)/Trace(K)$ at large iteration number
(low regularization strength) in Fig. \ref{fig:iterbertero}b). 

The optimum 
regularization strength and thus the final solution $S$ was determined
using the criterion  $Trace(T)/Trace(K)=1$. The estimates $\sum_j{M_{ij}\cdot S_j}$
(open circles) of $Y_i$ are compared with $Y_i$ (histogram) in Fig. \ref{fig:CrabData}b). 
These data enter in the calculation of $\chi_0^2$ (eq. \ref{eq:chi20}). The number of 
degrees of freedom in the unfolding procedure 
is $na-nb+1=5$, because the number of measurements is $na$, 
the number of unknowns is $nb$, and the relation $\sum_j S_j=\sum_i Y_i$ 
(eq. \ref{eq:addconstraint}) is used 
as additional constraint. As can be seen from Figs. \ref{fig:iterbertero}d)
and e) the values of $Reg(S)_{Tikhonov}$ and $Reg(S)_{Schmelling}$ are much lower at the 
selected regularization strength than without regularization. This means that
the solution $S$ is smoothed by the regularization. The black symbols in 
Fig. \ref{fig:CrabResults}a) represent the final solution $S$. The solution before correctiong 
for the cut in $E_{est}$ is drawn with grey symbols. \\   

The final differential $\gamma$-ray flux $\Phi$, as computed from $S$ 
according to eq.(\ref{eq:Phik}),
is drawn in Fig. \ref{fig:CrabResults}b). The solid line represents the result of a fit of
the expression 
$f_0\cdot \left(\dfrac{E}{300\;GeV}\right)^{\alpha}$,  
with $\alpha=a+b\cdot\log_{10}\left(\dfrac{E}{300\;GeV}\right)$,
to the data points.
The number of degrees of freedom for this fit is $nb-3= 7$, because the number of 
data points is $nb=10$ and the number of free parameters is 3 ($f_0$, $a$ and $b$).
The $\chi^2$ is 8 for 7 degrees of freedom. Setting $b=0$  in the fit yields
a  $\chi^2$ of 24 for 8 degrees of freedom. This fit is clearly disfavoured
as compared to the fit in which the slope $\alpha$ is energy dependent.
In these fits the full correlation matrix $T$ of $S$ has been taken into account.

The result of the latter fit for $\Phi(E_{true})$ was used to recalculate
the averages  $\overline{M}_i$ (\ref{eq60}) and  $\overline{A}$ 
(\ref{eq44x}) for the individual $E_{true}$ bins. The unfolding was repeated
using the recalculated  $\overline{M}_i$ and  $\overline{A}$, yielding new results
for $S$, $\Phi$ and the fit parameters. After 1 iteration this procedure converged.

Very similar results were obtained with the other unfolding methods. The spread of 
the $S_j$, obtained with the different unfolding methods, can be seen
in Fig. \ref{fig:CrabResults}c). This spread can be regarded as an estimate of
the a systematic error due to the unfolding.

\section{Correction Factors}
\label{section:correctionfactors}
A widely used method of correcting experimental distributions is the
application of correction factors: Using Monte Carlo data, both the
'true' distribution $S_k^{MC}\;\;\;\;(k=1, ...nc)$ and the 'reconstructed' 
distribution $Y_k^{MC}\;\;\;\;(k=1, ...nc)$ of some quantity are
produced under certain conditions (selections, cuts). Correction
factors are determined according to  
\begin{eqnarray}
c_k\;=\;S_k^{MC}/Y_k^{MC}
\qquad\qquad\qquad(k=1, ...nc)
\label{eq27}
\end{eqnarray}
An experimental distribution $Y_k$, obtained under the same conditions
as $Y_k^{MC}$, is then corrected by
\begin{eqnarray}
S_k\;=\;Y_k\cdot c_k
\qquad\qquad\qquad(k=1, ...nc)
\label{eq28}
\end{eqnarray}
to obtain the corrected distribution $S_k$.

The following properties of this procedure can be stated
\cite{schmelling98} :
\begin{itemize}
\item $c_k$ is undefined if $Y_k^{MC}=0$.
\item If $S_k^{MC}=0$ also $c_k=0$ and $S_k=0$, which means that $Y_k$
is ignored.
\item $c_k$ depends on the shape of the MC distribution $S_k^{MC}$; the
corrected distribution $S_k$ is always biased towards $S_k^{MC}$.
\item If $Y_k=0$ also $S_k$ is zero.
\item The standard linear error propagation often yields too small
errors of $S_k$.
\end{itemize}
The correction factors are only right if $S_k^{MC}$ is identical to
the true $S_k$ distribution. If this is not the case one may iterate
$S_k^{MC}$, setting $S_k^{MC}$ equal to the last corrected
experimental distribution $S_k$. However, this often leads to
instabilities. The reason for the instabilities appears to be similar
to that causing a large noise component of the direct solution (\ref{eq:S0}).

In contrast to the unfolding methods presented in Section \ref{section:regularization}, 
there is very little freedom in
choosing the binnings for $S$ and $Y$. By definition, the range of
values and the binnings for the true and reconstructed quantity are identical.

Advantages of the method of correction factors are that it is simple
and stable. The drawbacks have been listed above, the severest one
being the strong dependence of the
correction factors on the assumptions made in the MC about $S$.

\section{Summary}
\label{section:summary}
In this paper the procedures to unfold experimental energy distributions of
$\gamma$-rays, as applied in the MAGIC experiment, are described. It is explained, 
how the uncertainties, which are inherent in any unfolding process, can be
handled successfully. Possible problems in the unfolding are discussed and
suggestions are given which can help to avoid them. Various techniques are 
presented, which allow to reconstruct the energy spectrum in a rather unbiased way.
All algorithms are impleneted in the MAGIC software, which is based on the 
$C^{++}$ language and ROOT
\cite{ROOT}. Their application to real data has shown to provide robust and 
reliable results. The methods and procedures are applied in most of the MAGIC
analyses.



\appendix

\section{Acknowledgements}
We thank Michael Schmelling for fruitful discussions and critical comments.






\begin{thebibliography}{10}
\bibitem{MAGIC-commissioning}
Baixeras, C. et~al.,
\newblock 2004, Nucl. Instrum. Meth.,  A518, 188.
\bibitem{gold64}R.~Gold, ANL-6984 (1964)
\bibitem{marchuk75}G.I.~Marchuk, "Methods of Numerical Mathematics",
Springer, Berlin (1975)
\bibitem{tikhonov79}A.N.~Tikhonov and V.Ja.~Arsenin, Methods of
Solution of Ill-posed Problems - M (Nauka, 1979)
\bibitem{provencher82}S.W.~Provencher, Computer Physics Communications
27 (1982) 213 and 229
\bibitem{blobel84}V.~Blobel, "Unfolding methods in high-energy
physics experiments", DESY 84-118 (1984).
\bibitem{blobel84a}V.~Blobel, 1984 CERN School of Computing,
Ajguablava, Spain, CERN 85-09 (1984) p.88.
\bibitem{belogorlov85}E.A.~Belogorlov et al., "Interpretation of the
solution to the inverse problem for the positive function and the
reconstruction of neutron spectra", NIM A 235 (1985) 146.
\bibitem{giljazov87}S.F.~Giljazov, "Methods of Solution of Linear
Ill-posed Problems", MSU, Moscow (1987)
\bibitem{zhigunov88}V.P.~Zhigunov et al., "On estimating
distributions with the maximum entropy principle", NIM A 273 (1988) 362
\bibitem{bertero88}M.~Bertero, INFN/TC-88/2 (1988).
\bibitem{bertero89}M.~Bertero, Advances in Electronics and Electron
Physics, Vol.75 (1989).
\bibitem{anykeyev91}V.B.~Anykeyev et al., NIM A 303 (1991) 350.
\bibitem{schmelling94}M.~Schmelling, "The method of reduced
cross-entropy. A general approach to unfold probability
distributions", NIM A 340 (1994) 400.
\bibitem{blobel96}V.~Blobel, "The RUN manual", OPAL Technical Note
TN361 (1996)
\bibitem{hoecker96}A.~H\"ocker and V.~Kartvelishvili, NIM A 372 (1996) 469.
\bibitem{schmelling98}M.~Schmelling, "Numerische Methoden der
Datenanalyse", MPI-K Heidelberg (1998)
\bibitem{blobel02}V.~Blobel, "An Unfolding Method for High Energy Physics",
IPPP Workshop on Advanced Statistics Techniques in Particle Physics, Durham (2002)
\bibitem{wittek99}W.~Wittek,"Correlations between Parameters of
Extended Air Showers and their Proper Use in Analyses", 26th
Int. Cosmic Ray Conference, Salt Lake City, Utah, USA (1999)
\bibitem{Majumdar2005}
Majumdar, P. et~al. (MAGIC Collab.),
\newblock 2005, Proc. of the 29th ICRC, Pune, India, 5-203,
astro-ph/0508274.
\bibitem{Mizobuchi2005}
Mizobuchi, S. et~al. (MAGIC Collab.),
\newblock Proc. 29th ICRC, Pune, India, 5-323, astro-ph/0508274.
\bibitem{MARS}
Bretz, T. and Wagner, R. (MAGIC Collab.),
\newblock 2003, Proceedings of the 28$^{\mathrm{th}}$ ICRC, Tsukuba,
Japan, 2947.
\bibitem{kneiske04}T.M.~Kneiske et al., A\&A 413 (2004) 807
\bibitem{albert07}J.~Albert et al., submitted for publication in ApJ (2007), arXiv:0705.3244.
\bibitem{ROOT}R.~Brun, F.~Rademakers, http://root.cern.ch/.
\end{thebibliography}
\end{document}